\documentclass[3p,times,procedia]{elsarticle}
\usepackage{nupha_ecrc}
\usepackage{amsmath}

\volume{00}

\firstpage{1}

\journalname{Nuclear Physics A}

\runauth{}

\jid{nupha}

\jnltitlelogo{Nuclear Physics A}

\usepackage{amssymb}

\usepackage[figuresright]{rotating}

\newcommand{\xt}{\mathbf{x}}

\begin{document}

\begin{frontmatter}



\dochead{XXVIth International Conference on Ultrarelativistic Nucleus-Nucleus Collisions\\ (Quark Matter 2017)}

\title{3-D Glasma initial state from small-x evolution}


\author[BS]{Bj\"{o}rn Schenke}
\author[SS]{S\"{o}ren Schlichting}

\address[BS]{Physics Department, Brookhaven National Laboratory, Bldg. 510A, Upton, NY 11973, USA}
\address[SS]{Department of Physics, University of Washington, Seattle, WA 98195-1560, USA}

\address{}

\begin{abstract}
We present an ab-initio approach to compute the longitudinal dependence of the initial state in heavy-ion collisions by including small-x evolution of the nuclear gluon distributions. Extending the IP-Glasma model by including JIMWLK rapidity evolution, we compute event-by-event rapidity distributions of produced gluons and the early time energy momentum tensor as a function of space-time rapidity and transverse coordinates. We show how the effects of small-x evolution manifest themselves in longitudinal (rapidity) correlations of event-by-event multiplicities and transverse geometry and compare our results to various phenomenological models and experimental observations.
\end{abstract}


\end{frontmatter}


\section{Introduction}
\label{seq:Intro}
Experimental measurements of longitudinal correlations and fluctuations \cite{Khachatryan:2015oea,ATLAS-CONF-2015-020,Aaboud:2016jnr,ATLAS-CONF-2017-003} have lead to interesting new insights into the dynamics of multi-particle production and the 3+1-D space-time dynamics of the QCD medium created in high-energy heavy-ion collisions. On the phenomenological side, a variety of different models~\cite{Steinheimer:2007iy,Bozek:2015bna,Pang:2015zrq,Monnai:2015sca} have been developed in an attempt to understand the characteristic behavior of observables, such as longitudinal multiplicity distributions and event plane correlations. Despite the fact that different models perform reasonably well in comparison with experimental results, the underlying dynamics appears to be far from understood as different models are based on vastly different degrees of freedom ranging from constituent quarks~\cite{Monnai:2015sca} and flux tubes/strings~\cite{Werner:2010aa,Pang:2015zrq,Bozek:2015bna} all the way to hadronic degrees of freedom~\cite{Steinheimer:2007iy}. In this proceeding, we report on a first attempt to understand the structure of longitudinal correlations and fluctuations from a first-principle approach in the high-energy limit. Based on a brief exposition of the formalism, we present first results for longitudinal observables in central $Pb+Pb$ collisions at LHC energies and give an outlook to possible future extensions. Details of our study, including a comprehensive discussion of the theoretical background as well as some additional results can be found in~\cite{Schenke:2016ksl}.
\section{3-D initial state from small-x evolution}
\label{seq:Main}
Energy deposition at mid-rapidity is dominated by small-$x$ gluons in the wave-functions of the incoming nuclei and admits an effective description in the Color Glass Condensate framework~\cite{Iancu:2003xm,Gelis:2010nm} in the high-energy limit.  Based on the property of high-energy factorization, expectation values of single inclusive observables, such as e.g. the average multiplicity $\left.dN/dy\right|_{y_{\rm obs}}$, can be computed to leading logarithmic accuracy in this framework as an average over the distribution of color charges in the projectile and target according to~\cite{Gelis:2008rw,Gelis:2008ad,Gelis:2008sz}
\begin{eqnarray}
\label{eq:fac}
\left.\frac{dN}{dy}\right|_{y_{\rm obs}} = \int [DU] [DV]~\mathcal{W}^{p}_{\Delta y_p}[U]~\mathcal{W}^{t}_{\Delta y_t}[V]~\frac{dN}{dy}[U,V]\;.
\end{eqnarray}
where the evolution of the weight functionals $\mathcal{W}^{p/t}_{\Delta y_{p/t}}$ of the projectile/target with the rapidity separation $\Delta y_{p/t}=\pm y_{p/t}-y_{\rm obs}$ is described by the JIMWLK renormalization group equation~\cite{JalilianMarian:1996xn,JalilianMarian:1997jx,JalilianMarian:1997gr,Iancu:2000hn,Iancu:2001ad}. We note that high-energy factorization in the above form ~(\ref{eq:fac}) has been proven in~\cite{Gelis:2008rw,Gelis:2008ad,Gelis:2008sz} for single inclusive observables, e.g $\left.dN/dy\right|_{y_{\rm obs}}$  and un-equal rapidity correlations, e.g. $\left.dN/dy\right|_{y_{\rm obs}^{1}} \left.dN/dy\right|_{y_{\rm obs}^{2}}$, with sufficiently small rapidity separations $|y_{\rm obs}^{1}-y_{\rm obs}^{2}|\ll 1/\alpha_s$. Even though factorization of the form (\ref{eq:fac}) is expected to break down in the regime where $|y_{\rm obs}^{1}-y_{\rm obs}^{2}|\sim1/\alpha_s$ (see e.g.~\cite{Iancu:2013uva}), it is conceivable that even in this regime the above expression still contains the dominant source of correlations, and we will assume in the following that the factorization holds over the entire range of $|y_{\rm obs}^{1}-y_{\rm obs}^{2}|$ considered in our study.

Based on these general ideas, the IP-Glasma model~\cite{Schenke:2012wb,Schenke:2012hg,Gale:2012rq} represents a successful microscopic model for the phenomenological description of (2+1-D) transverse properties of the initial state. By use of the high-energy factorization properties, it is then straightforward to include the dominant sources of longitudinal correlations and fluctuations in a 3+1 D description of the initial state on an event-by-event basis. We first generate a configuration of Wilson lines $U/V$ according to the statistical weights $\mathcal{W}_{\pm y_{p/t}-y^{\rm init}_{p/t}}[U/V]$ of the IP-Sat paramatrization evaluated at an initial rapidity separation $y^{\rm init}_{p/t}$, corresponding to the largest forward/backward rapidity of interest ($y^{\rm init}_{p/t}=\pm2.4$). We then evolve the Wilson lines $U/V$ from the initial rapidity scale $\pm y_{p/t}-y^{\rm init}_{p/t}$ up to the rapidity separation $\pm y_{p/t}-y_{\rm obs}$ of interest, by solving the stochastic version of the leading log (LL) JIMWLK evolution equation~\cite{Weigert:2000gi,Blaizot:2002np}. \footnote{We note that even though NLL effects of the evolution are know to be important, e.g. to slow down the evolution of DIS structure functions~\cite{Iancu:2015joa}, we stay at LL accuracy and instead use the coupling constants $\alpha_s$ as a free parameter to adjust evolution speed. Since we are interested in impact parameter dependence, one has to introduce an infrared regulator to suppress gluon radiation at large distance scales~\cite{Schlichting:2014ipa}. Details of the procedure are discussed in~\cite{Schenke:2016ksl} where the sensitivity to the infrared regulator is also assessed.} Based on the Wilson line configurations  $\left.U/V\right|_{\Delta y_{p/t}}$ we then proceed as usual to compute the observables of interest by numerically solving the classical Yang-Mills equations of motion. Un-equal rapidity correlations such as e.g. the multiplicity correlator $\left\langle \left.dN/dy\right|_{y_{1}} \left.dN/dy\right|_{y_{2}} \right\rangle$ are simply obtained as event-averages of the product of the local observables $\left.dN/dy\right|_{y_{1}}$ and $\left.dN/dy\right|_{y_{2}}$ and reflect the underlying correlations of small-x gluons inside the hadronic wave-functions.

\begin{figure}[t!]
\centering
\begin{minipage}{0.6\textwidth}
\includegraphics[width=0.9\textwidth]{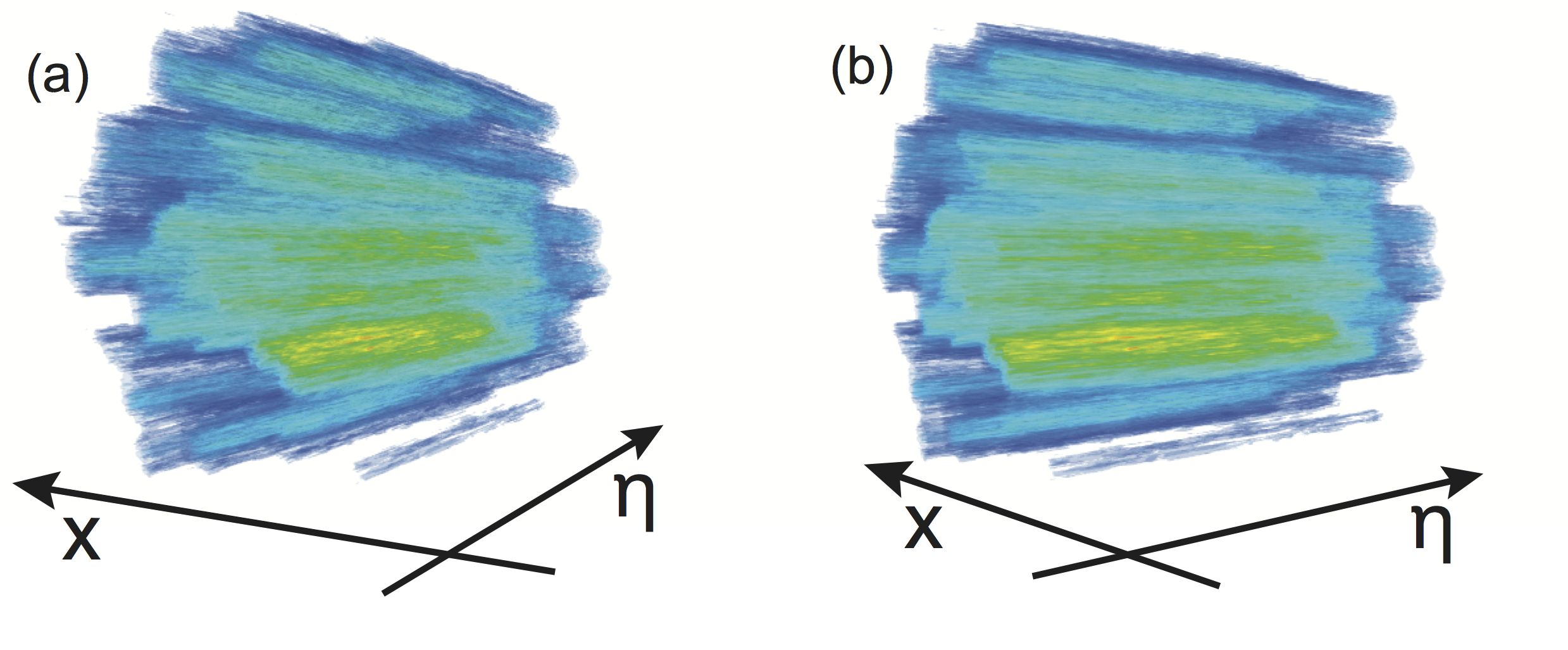}%
\end{minipage}
\begin{minipage}{0.39\textwidth}
\includegraphics[width=0.9\textwidth]{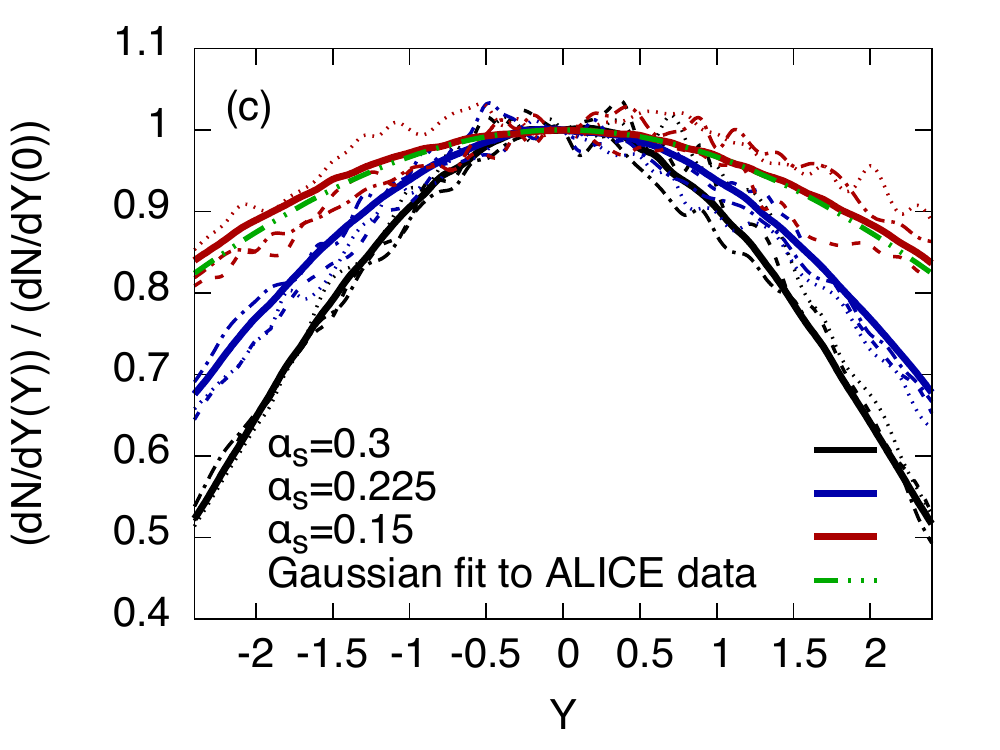}%
\end{minipage}
\caption{\label{fig:EventProfile}(a-b) Illustration of 3-D energy density profile in a single $Pb+Pb$ event. (c) Event averaged multiplicity distribution $\frac{dN}{dy}$. Fig. from~\cite{Schenke:2016ksl}}
\end{figure}
Before we turn to a detailed discussion of our results in comparison to experimental data, we briefly comment on the general features of our 3-D Glasma initial state model. So far all of our calculations are based on initial state observables only, i.e., do not include hydrodynamic evolution, and have only been performed for 2.76 Pb+Pb collisions at zero impact parameter $(b=0)$. In the left panel of Fig.~\ref{fig:EventProfile} we present a contour plot of the initial state energy density $(T^{\tau\tau})$ in a single $Pb+Pb$ event. One clearly observes that energy deposition is dominated by approximately boost invariant flux tubes with a characteristic transverse size on the order of the size of the nucleon. However, one also observes various kinds of short and long range fluctuations both in the longitudinal ($\eta$) and transverse ($x$) directions, which can be quantified further in terms of various correlation functions.

Before we turn to a more detailed analysis of the longitudinal correlations and fluctuations, it is useful to check that our model correctly reproduces the rapidity dependence of the single inclusive multiplicity distributions. Our results for $dN/dy$ (normalized to $dN/dy$ at mid-rapdidity) are shown in the right panel of Fig.~\ref{fig:EventProfile}. Since  the rapidity evolution speed is controlled by the strong coupling constant $\alpha_s$, we present results for three different values of $\alpha_s$ noting that the distributions become almost indistinguishable from each other when plotted as a function of the scaling variable $\alpha_s Y$. Comparing our results with a Gaussian fit to ALICE data~\cite{Abbas:2013bpa}, one observes good overall agreement for $\alpha_s=0.15$. Even though, this value of the strong coupling constant appears to be somewhat small, it is re-assuring that similar values of $\alpha_s$ are needed to match the evolution speeds of the saturation scale $d \ln Q_s^2/dy\simeq 0.3$ with phenomenological extractions of $Q_s^2$ from DIS structure functions~\cite{GolecBiernat:1998js,Iancu:2003ge,Rezaeian:2013tka}.

We have also analyzed longitudinal multiplicity fluctuations, characterized by the unequal-rapidity correlation function $C(Y_1,Y_2) \propto \langle dN/dY_1 dN/dY_2 \rangle/   \langle dN/dY_1 \rangle \langle dN/dY_2 \rangle$. We find that our model~\cite{Schenke:2016ksl} correctly reproduces the overall structure of the correlation function measured by ATLAS~\cite{ATLAS-CONF-2015-020}, where multiplicity fluctuations are dominated by an event-by-event forward/backward asymmetry. We note that in our framework the dominant contribution to this forward/backward asymmetry comes from local fluctuations of the saturation scale $Q_s$ at the initial rapidity scale, and we refer to~\cite{Bzdak:2015eii} for a semi-analytic discussion within a closely related approach.

\begin{figure}[t!]
\centering
\begin{minipage}{0.45\textwidth}
\includegraphics[width=0.9\textwidth]{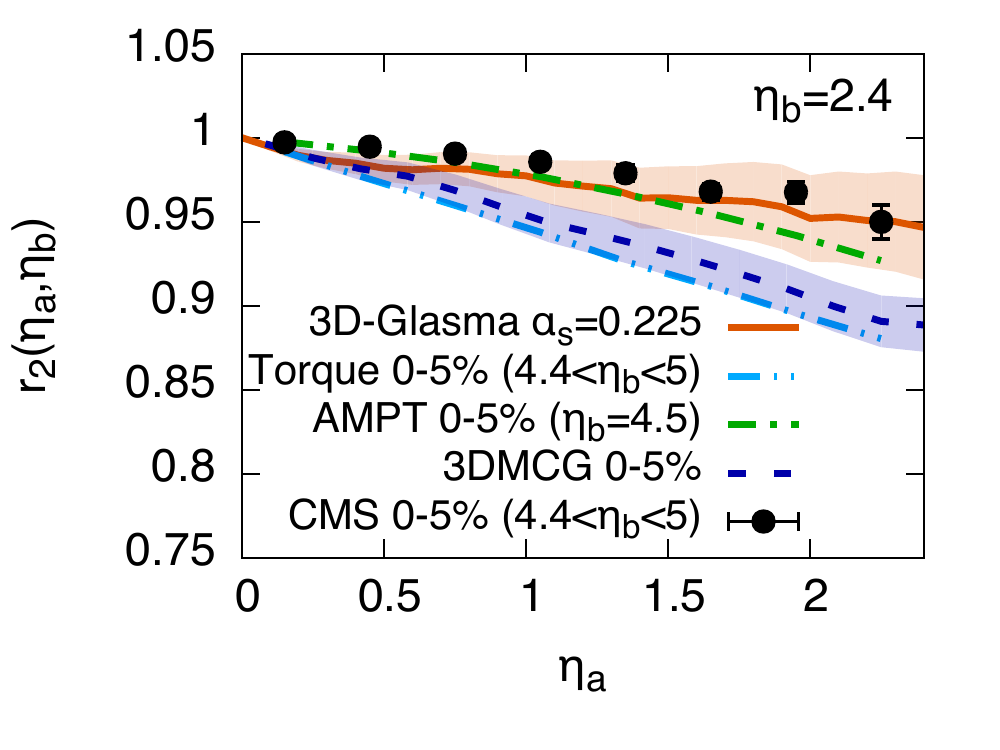}%
\end{minipage}
\begin{minipage}{0.45\textwidth}
\includegraphics[width=0.9\textwidth]{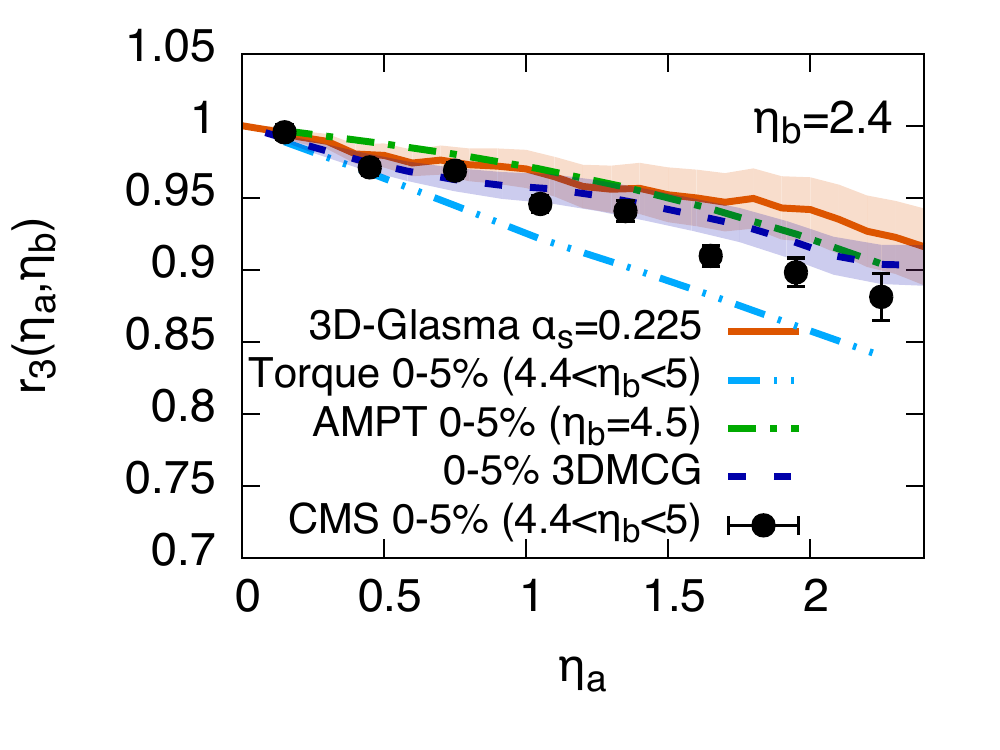}%
\end{minipage}
\caption{\label{fig:r2r3} De-correlation of the elliptic ($r_{2}$) and triangular ($r_{3}$) event-geometry as a function of rapidity separation $\eta_a$ (see text for details). Different calculations (3D-Glasma~\cite{Schenke:2016ksl},~Torque~\cite{Bozek:2015bna},~AMPT~\cite{Pang:2015zrq},~3DMCG~\cite{Monnai:2015sca}) are compared to CMS results~\cite{Khachatryan:2015oea}. Fig. from~\cite{Schenke:2016ksl}.}
\end{figure}

We finally address the longitudinal fluctuations of the transverse event geometry within the 3-D Glasma model. Characterizing the initial state geometry in terms of the usual eccentricity vectors
$\boldsymbol{\epsilon}_{n}(\eta)=\int d^2\xt~|\xt|^n$ $~e^{in\phi_{n}(\xt)}~T^{\tau\tau}(\xt,\eta)~/~\int d^2\xt~|\xt|^nT^{\tau\tau}(\xt,\eta),$
we observe a significant de-correlation of both the magnitude of the eccentricity $|\boldsymbol{\epsilon}_{n}(\eta)|$ and the eccentricity plane angles $\psi_{n}(\eta)=\arg(\boldsymbol{\epsilon}_{n}(\eta))/n$ for both elliptic $(n=2)$ and triangular $(n=3)$ eccentricities across the entire rapidity range of $-2.4 < \eta < 2.4$. We can further quantify this de-correlation by studying the forward/backward ratio~\cite{Khachatryan:2015oea}, $r_{n}(\eta_a,\eta_b)= Re\Big[ \boldsymbol{\epsilon}_{n}(+\eta_a) \boldsymbol{\epsilon}^{*}_{n}(\eta_b) \Big] / Re\Big[ \boldsymbol{\epsilon}_{n}(-\eta_a) \boldsymbol{\epsilon}^{*}_{n}(\eta_b) \Big]$ measured w.r.t. a reference rapidity $\eta_b$. Since previous works~\cite{Pang:2015zrq} showed that the magnitude of the initial state $r_{n}$ defined on the basis of eccentricities is close to that of the final state $r_{n}$, where the de-correlation ratio is calculated based on the usual flow vectors $Q_{n}$, we can directly compare our results to experimental measurements. Our results for $r_2$ and $r_3$ are compactly summarized in Fig.~\ref{fig:r2r3}, where we present a comparison between various models~\cite{Schenke:2016ksl,Bozek:2015bna,Pang:2015zrq,Monnai:2015sca} and experimental results from CMS for central (0-5\%) $Pb+Pb$ collisions~\cite{Khachatryan:2015oea}. Despite the fact that all models predict a significant de-correlation $(r_{2/3}<1)$, a simultaneous description of $r_2$ and $r_3$ appears to be challenging as models typically over-estimate the de-correlation of $r_2$ in comparison to $r_3$. We find that with the 3D-Glasma model we obtain a decent description of both $r_2$ and $r_3$, although more detailed studies will be needed to reduce statistical and systematic errors in this comparison. 

\section{Conclusions \& Outlook}
\label{seq:Conlusions}
We reported on the development of a 3-D initial state model based on the concept of high-energy factorization. One particularly appealing feature of this approach is that the longitudinal structure of the event geometry is entirely determined by the small-x (JIMLWK) evolution, such that the strong coupling constant $\alpha_s$ and a non-perturbative infrared regulator $m$ (see~\cite{Schenke:2016ksl} for details) are the only free parameters of the model.  We presented first applications of the 3-D Glasma model to the initial state in central $Pb+Pb$ collisions, which show promising results concerning the description of the rapidity dependence of the avg. multiplicity as well as the long. structure of multiplicity and geometric correlations. By including running coupling effects and performing proper event selection, it will be interesting to extend this study towards a more comprehensive analysis of the longitudinal structure in different collision systems $(p+p,p+A,A+A)$.

\textit{Acknowledgements:} We acknowledge support by the U.S. Department of Energy under Grant No. DE-SC0012704 (B.S.) and DE-FG02-97ER41014 (S.S.). Numerical calculations used resources of NERSC, a DOE Office of Science User Facility supported by the Office of Science of the U.S. Department of Energy under Contract No. DE-AC02-05CH11231. 





\bibliographystyle{elsarticle-num}

\begin{thebibliography}{99}

\bibitem{Khachatryan:2015oea} 
  V.~Khachatryan {\it et al.} [CMS Collaboration],
  Phys.\ Rev.\ C {\bf 92}, no. 3, 034911 (2015)


\bibitem{ATLAS-CONF-2015-020} 
  The ATLAS collaboration [ATLAS Collaboration],
  ATLAS-CONF-2015-020.


\bibitem{Aaboud:2016jnr} 
  M.~Aaboud {\it et al.} [ATLAS Collaboration],
  arXiv:1606.08170 [hep-ex].


\bibitem{ATLAS-CONF-2017-003} 
  The ATLAS collaboration [ATLAS Collaboration],
  ATLAS-CONF-2017-003.


\bibitem{Steinheimer:2007iy} 
  J.~Steinheimer, M.~Bleicher, H.~Petersen, S.~Schramm, H.~Stocker and D.~Zschiesche,
  Phys.\ Rev.\ C {\bf 77}, 034901 (2008)


\bibitem{Bozek:2015bna} 
  P.~Bozek and W.~Broniowski,
  Phys.\ Lett.\ B {\bf 752}, 206 (2016)


\bibitem{Pang:2015zrq} 
  L.~G.~Pang, H.~Petersen, G.~Y.~Qin, V.~Roy and X.~N.~Wang,
  Eur.\ Phys.\ J.\ A {\bf 52}, no. 4, 97 (2016)


\bibitem{Monnai:2015sca} 
  A.~Monnai and B.~Schenke,
  Phys.\ Lett.\ B {\bf 752}, 317 (2016)


\bibitem{Werner:2010aa} 
  K.~Werner, I.~Karpenko, T.~Pierog, M.~Bleicher and K.~Mikhailov,
  Phys.\ Rev.\ C {\bf 82}, 044904 (2010)


\bibitem{Schenke:2016ksl} 
  B.~Schenke and S.~Schlichting,
  Phys.\ Rev.\ C {\bf 94}, no. 4, 044907 (2016)


\bibitem{Iancu:2003xm} 
  E.~Iancu and R.~Venugopalan,
  In *Hwa, R.C. (ed.) et al.: Quark gluon plasma* 249-3363


\bibitem{Gelis:2010nm} 
  F.~Gelis, E.~Iancu, J.~Jalilian-Marian and R.~Venugopalan,
  Ann.\ Rev.\ Nucl.\ Part.\ Sci.\  {\bf 60}, 463 (2010)


\bibitem{Gelis:2008rw} 
  F.~Gelis, T.~Lappi and R.~Venugopalan,
  Phys.\ Rev.\ D {\bf 78}, 054019 (2008)


\bibitem{Gelis:2008ad} 
  F.~Gelis, T.~Lappi and R.~Venugopalan,
  Phys.\ Rev.\ D {\bf 78}, 054020 (2008)


\bibitem{Gelis:2008sz} 
  F.~Gelis, T.~Lappi and R.~Venugopalan,
  Phys.\ Rev.\ D {\bf 79}, 094017 (2009)


\bibitem{JalilianMarian:1996xn} 
  J.~Jalilian-Marian, A.~Kovner, L.~D.~McLerran and H.~Weigert,
  Phys.\ Rev.\ D {\bf 55}, 5414 (1997)


\bibitem{JalilianMarian:1997jx} 
  J.~Jalilian-Marian, A.~Kovner, A.~Leonidov and H.~Weigert,
  Nucl.\ Phys.\ B {\bf 504}, 415 (1997)


\bibitem{JalilianMarian:1997gr} 
  J.~Jalilian-Marian, A.~Kovner, A.~Leonidov and H.~Weigert,
  Phys.\ Rev.\ D {\bf 59}, 014014 (1998)


\bibitem{Iancu:2000hn} 
  E.~Iancu, A.~Leonidov and L.~D.~McLerran,
  Nucl.\ Phys.\ A {\bf 692}, 583 (2001)


\bibitem{Iancu:2001ad} 
  E.~Iancu, A.~Leonidov and L.~D.~McLerran,
  Phys.\ Lett.\ B {\bf 510}, 133 (2001)


\bibitem{Iancu:2013uva} 
  E.~Iancu and D.~N.~Triantafyllopoulos,
  JHEP {\bf 1311}, 067 (2013)


\bibitem{Schenke:2012wb} 
  B.~Schenke, P.~Tribedy and R.~Venugopalan,
  Phys.\ Rev.\ Lett.\  {\bf 108}, 252301 (2012)


\bibitem{Schenke:2012hg} 
  B.~Schenke, P.~Tribedy and R.~Venugopalan,
  Phys.\ Rev.\ C {\bf 86}, 034908 (2012)


\bibitem{Gale:2012rq} 
  C.~Gale, S.~Jeon, B.~Schenke, P.~Tribedy and R.~Venugopalan,
  Phys.\ Rev.\ Lett.\  {\bf 110}, no. 1, 012302 (2013)


\bibitem{Weigert:2000gi} 
  H.~Weigert,
  Nucl.\ Phys.\ A {\bf 703}, 823 (2002)


\bibitem{Blaizot:2002np} 
  J.~P.~Blaizot, E.~Iancu and H.~Weigert,
  Nucl.\ Phys.\ A {\bf 713}, 441 (2003)


\bibitem{Iancu:2015joa} 
  E.~Iancu, J.~D.~Madrigal, A.~H.~Mueller, G.~Soyez and D.~N.~Triantafyllopoulos,
  Phys.\ Lett.\ B {\bf 750}, 643 (2015)


\bibitem{Schlichting:2014ipa} 
  S.~Schlichting and B.~Schenke,
  Phys.\ Lett.\ B {\bf 739}, 313 (2014)


\bibitem{Abbas:2013bpa} 
  E.~Abbas {\it et al.} [ALICE Collaboration],
  Phys.\ Lett.\ B {\bf 726}, 610 (2013)


\bibitem{GolecBiernat:1998js} 
  K.~J.~Golec-Biernat and M.~Wusthoff,
  Phys.\ Rev.\ D {\bf 59}, 014017 (1998)


\bibitem{Iancu:2003ge} 
  E.~Iancu, K.~Itakura and S.~Munier,
  Phys.\ Lett.\ B {\bf 590}, 199 (2004)


\bibitem{Rezaeian:2013tka} 
  A.~H.~Rezaeian and I.~Schmidt,
  Phys.\ Rev.\ D {\bf 88}, 074016 (2013)


\bibitem{Bzdak:2015eii} 
  A.~Bzdak and K.~Dusling,
  Phys.\ Rev.\ C {\bf 93}, no. 3, 031901 (2016)

\end{thebibliography}



\end{document}